\documentclass[seceq]{ptptex}

\usepackage{graphicx}
\usepackage{wrapft}

\title{
Analysis of the Phase Structure of Thermal QED/QCD through \\
the HTL Improved Ladder Dyson-Schwinger Equation 
}
\subtitle{On the Gauge Dependence of the Solution}    

\author{
Hisao {\sc Nakkagawa},\footnote{E-mail: 
nakk@daibutsu.nara-u.ac.jp}
Hiroshi {\sc Yokota}\footnote{E-mail:
yokotah@daibutsu.nara-u.ac.jp}
and Koji {\sc Yoshida}\footnote{E-mail: 
yoshidak@daibutsu.nara-u.ac.jp}
}

\inst{
Institute for  Natural Sciences,  Nara University, Nara 631-8502, Japan
}

\abst{
We solved with a numerical procedure the HTL improved ladder DS equation for the retarded fermion self-energy 
function $\Sigma_R$ to study the spontaneous generation of fermion mass in thermal QCD/QED, and studied the 
gauge-dependence of the solution within a general covariant gauge where the gauge parameter $\xi$ is any 
constant number.

With the numerical solutions thus obtained, we found the followings; i) The fermion wave function renormalization 
function $A(P)$ always deviates largely from unity even at the momentum where the mass is defined, thus the 
corresponding solutions explicitly contradict with the Ward-Takahashi identity.  ii) As a result, the obtained 
solutions strongly depend on the choice of gauge parameters: the critical temperatures and the critical coupling 
constants significantly change gauge by gauge. In all gauges we studied in the present analysis, we could not find 
any solution, having a possibility to be consistent with the Ward-Takahashi identity. Thus we are forced to 
investigate the procedure to find a gauge which enables us to get a solution being consistent with the Ward-Takahashi 
identity, otherwise we can not obtain any physically sensible conclusions through the analysis of the point-vertex 
ladder DS equation no matter how the gauge propagator gets improved.
}

\begin{document}

\maketitle

\newcounter{beanr}

\section{Introduction and summary}
\label{s1}

Recent studies on gauge field theory have been revealing rich aspects of phase structure of the matter according to
the variation of number density and/or temperature. However, it is supposed to be hard to obtain more detailed 
understanding of the mechanism of phase transition by means of such theoretical studies, since most of them were 
carried out on the basis of perturbative calculation or of numerical lattice simulation. This motivated us to survey 
the problem of phase transition using the Dyson-Schwinger (DS) equation, firstly because the DS equation is derived exactly 
in the field theory and is the fundamental equation to investigate nonperturbative phenomena within the framework 
of field theory, and secondly because we can obtain successively improved solutions by the successive refinement of the 
analytical approximation to its integration kernel, thus revealing the essential contribution that controls the phase
transition depending on the temperature/density. 

We started our analysis with the DS equation for the retarded fermion mass function  $\Sigma_R$ to study the 
spontaneous generation of fermion mass in thermal QED/QCD.\cite{FNYY02} In the analysis we used an improved ladder interaction 
kernel obtained analytically through the Hard-Thermal-Loop (HTL) resummation procedure, in which the ladder kernel 
is improved by use of the HTL resummed form of the gauge boson propagator. We realized that the results obtained\cite{FNYY03}
are significantly different from those obtained in the preceding analyses with the simple ladder DS equation.\cite{Barducci}
It is worth noticing that, in all preceding DS equation analyses including ours, the bare (point) vertex ladder 
approximation for the integration kernel has been used, thus no result considering any vertex correction.

We here summarize the essential points of our analysis that may give the result significantly different from those of 
the preceding works. (For details of our results, see Ref.~\citen{FNYY03}.)
\begin{list}{\roman{beanr})}{\usecounter{beanr}}
  \item The DS equation for the retarded fermion mass function  $\Sigma_R$ is derived correctly without any specific 
        assumption for its form.
  \item  Nontrivial imaginary parts of the invariant functions $A$, $B$ and $C$ (see, the definition of $\Sigma_R$, 
        Eq.~(\ref{eq_sigma})) are taken into account.
  \item We adopted the improved ladder integration kernel analytically obtained in the HTL approximation.
\end{list}

Despite the improvements in taking into account correctly a) the (unstable) thermal quasiparticle nature of the fermion 
in the heat bath and b) the dominant effect of thermal fluctuations through the HTL resummation, it was suggested that 
the obtained result showed a serious problem, i.e., dependence on the choice of the gauge used in the analysis.
We can recognize it to see that the Ward-Takahashi identity $Z_1 = Z_2$ does not hold, where $Z_1$ and $Z_2$ denote 
the vertex and the wave-function renormalization constants, respectively. 

The retarded fermion mass function can be parameterized with the  three invariant functions $A$, $B$ and $C$ as
\begin{equation}
\label{eq_sigma}
   \Sigma_R(P) = (1-A(P)) p_i \gamma^i - B(P) \gamma^0 + C(P)
\end{equation}
The inverse of $A(P)$ at the momentum where the fermion mass is calculated is nothing but the wave function renormalization
constant $Z_2$. As noted above, the vertex renormalization constant $Z_1$ is exactly unity, $Z_1=1$, in our analysis 
because of the bare (point) vertex ladder approximation for the integration kernel, which is also the case with other
works carried out before. It follows that the Ward-Takahashi identity, the statement of gauge invariance, requires 
$Z_2=1$, namely, $A(P)=1$ at least at the momentum where the fermion mass is calculated. If $A\neq 1$ in the obtained 
results, it means that the results do not satisfy gauge invariance and hence there is little physical meaning in the 
results obtained.

For the vacuum ($T=0$) case, fortunately, $A(P)=1$ is verified to hold in the analysis in the Landau gauge.\cite{Maskawa} 
Therefore it is supposed that the analysis of the DS equation in the Landau gauge, apart from the reliability of the ladder
approximation, has some physical significance in the quantitative as well as qualitative sense.\cite{Yamawaki,Bardeen} 
At finite temperatures, however, there is no guarantee that $A=1$ holds even at the momentum where the fermion mass 
is calculated. In fact our analysis in the Landau gauge shows that $A$ largely deviates from 1 and becomes even complex 
number.\cite{FNYY03} Namely the analysis of DS equation in the ladder approximation at finite temperature/density 
QED/QCD is obviously inconsistent with gauge invariance. 

This gauge-dependence problem must be tackled seriously 
in order to draw a definite conclusion from our analysis of the HTL improved ladder DS equation, which indicated 
the importance to correctly take the dominant effect of thermal fluctuations into the integration kernel through 
the HTL resummation.

To see the problem more clearly we must clarify how sensitive the obtained results are to the choice of gauges. It will 
help us to study whether there exists a solution of the DS eqution in the ladder approximation that satisfies the 
Ward-Takahashi identity $Z_1=Z_2$ at finite temperature/density, and also to investigate, if such a solution exists, 
what is the real difference of it from the ``gauge-dependent'' solutions.

To answer the questions, in this paper we will solve the HTL improved ladder DS equation in a general covariant gauge 
and study the dependence of the solutions on the various choices of the gauge parameter, then investigate the possibility 
to obtain a solution consistent with the Ward-Takahashi identity.  We also make an improvement in estimating the 
numerical integration over the singular part of integration kernel in the present analysis.

We here present the results of our analysis; We find, within gauges where the gauge parameter $\xi$ is constant 
numbers, i) that the fermion wave function renormalization function $A(P)$ always deviates largely from unity even 
at the momentum where the mass is defined, thus that the corresponding solutions explicitly contradict with the 
Ward-Takahashi identity, and ii) that, as a result, the obtained solutions strongly depend on the choice of gauge 
parameters: the critical temperatures and the critical coupling constants significantly change gauge by gauge. 
In all gauges we study in the present analysis, we can not find any solution that has a possibility to be consistent 
with the Ward-Takahashi identity. Thus we are forced to investigate the procedure to find a gauge which enables us 
to get  a solution being consistent with the Ward-Takahashi identity, otherwise we can not obtain any physically 
sensible conclusions through the analysis of the point-vertex ladder DS eqution no matter how the gauge propagator 
gets improved.

This paper is organized as follows; In the next $\S$\ref{s2} we present the HTL resummed Dyson-Schwinger equation 
for the retarded fermion self-energy function $\Sigma_R$, and explain the improved ladder approximation adopted 
in the present analysis. Section 3 is devoted to presenting the solutions of the HTL improved ladder DS equation, 
revealing the large gauge-dependence between solutions. Conclusions of the present paper and the related discussion 
are given in the last $\S$\ref{s4}. In the Appendix we give some technical details in the numerical analysis in 
solving the DS equations.

\section{The HTL resummed improved ladder Dyson-Schwinger equation}
\label{s2}

In this section we present the HTL resummed DS equation for the retarded fermion self-energy function $\Sigma_R$. 
We also give an explication about the improved ladder approximation we make use of to the HTL resummed gauge 
boson propagator.

We then present the HTL resummed improved ladder Dyson-Schwinger equation for the independent invariant scalar 
functions $A$, $B$ and $C$. We also calculate the effective potential for the retarded fermion propagator $S_R$ 
in order to find the ``true solution'' when we get several ``solutions'' of the DS equation.  

\subsection{The HTL resummed Dyson-Schwinger equation for the retarded fermion self-energy function $\Sigma_R$}
\label{s21}

In the real time closed time-path formalism, we obtain, in the massless thermal QED/QCD in the HTL approximation, 
the DS equation for retarded fermion self-energy function $\Sigma_R$:
\begin{eqnarray}
\label{eq_DS} 
& & - i\Sigma_R (P)  = -i \Sigma_{RA}(-P,P) =
        - \frac{e^2}{2} \int \frac{d^4K}{(2\pi)^4} 
       \nonumber \\
  &  & \ \ \ \times \left[ ^*\Gamma^{\mu}_{RAA}(-P,K,P-K) S_{RA}(K) 
       \ ^*\Gamma^{\nu}_{RAA}(-K,P,K-P) \ ^*G_{RR,\mu\nu} (P-K) 
       \right. \nonumber \\
  & & \ \ \ \ \left. + \ ^*\Gamma^{\mu}_{RAA}(-P,K,P-K) S_{RR}(K)
       \ ^*\Gamma^{\nu}_{AAR}(-K,P,K-P) \ ^*G_{RA, \mu\nu}(P-K)
       \right]\ , \nonumber \\
\end{eqnarray}
Here $^*G^{\mu \nu}$ is the HTL resummed gauge boson propagator, whose  retarded ($R \equiv RA$) and 
correlation ($C \equiv RR$) components are given by\cite{Weldon} 
\begin{eqnarray}
\label{eq:Gr}
{}^*G_R^{\mu\nu}(K) \!\! &\equiv& \!\! {}^*G_{RA}^{\mu\nu} (-K,K) \nonumber \\
\!\! &=& \!\! \frac{1}{{}^*\Pi^R_T(K) -K^2 - i \epsilon k_0} A^{\mu \nu}
    + \frac{1}{{}^*\Pi^R_L(K) -K^2 - i \epsilon k_0} B^{\mu \nu}
    - \frac{\xi}{K^2 + i \epsilon k_0} D^{\mu \nu} , \nonumber \\
& & \\
{}^*G_C^{\mu\nu} (K) &\equiv& {}^*G_{RR}^{\mu\nu} (-K,K) \ = \ 
    (1+2n_B(k_0)) \left[ {}^*G_R^{\mu \nu}(K) - {}^*G_A^{\mu \nu}(K) \right] , \\
    n_B(k_0) &=& \frac{1}{\exp(k_0/T)-1} , 
\end{eqnarray}
with $^*\Pi^R_T$ and $^*\Pi^R_L$ being the HTL contributions to the transverse and longitudinal modes of the retarded 
gauge boson self-energy, respectively.\cite{Klimov} The parameter $\xi$ is the gauge-fixing parameter ($\xi=0$ in the Landau 
gauge). In the above, $A^{\mu \nu}$, $B^{\mu \nu}$ and $D^{\mu \nu}$ are the projection tensors given by\cite{Weldon}
\begin{equation}
A^{\mu\nu} = g^{\mu\nu}-B^{\mu\nu} -D^{\mu\nu}, \ \ 
B^{\mu \nu} = -\frac{\tilde{K}^{\mu} \tilde{K}^{\nu}}{K^2}, \ \ D_{\mu\nu}=\frac{K^{\mu} K^{\nu}}{K^2}, 
\end{equation}
where $\tilde{K}=(k,k_0 \hat{\bf k})$, $k=\sqrt{{\bf k}^2}$ and $\hat{\bf k}={\bf k}/k$ denotes the unit three 
vector along ${\bf k}$.

We use $S(-P,P) \equiv S(P)$ to denote the full fermion propagator, whose retarded ($R \equiv RA$) and 
correlation ($C \equiv RR$) components are given by 
\begin{eqnarray}
S_R (P) &\equiv& S_{RA} (-P,P) \ = \ \frac{1}{P\!\!\!\!/ + i \epsilon \gamma_0 - \Sigma_R} , \\
S_C (P) &\equiv& S_{RR} (-P,P) \ = \ (1-2n_F(p_0)) \left[ S_R(P) - S_A(P) \right] , \\
n_F(p_0) &=& \frac{1}{\exp(p_0/T)+1} ,
\end{eqnarray}
with the retarded fermion self-energy function $\Sigma_R$ decomposed as Eq.~(\ref{eq_sigma}) in terms of 
the independent invariant scalar functions $A(P)$, $B(P)$ and $C(P)$.

Finally, the HTL resummed 3-point fermion-gauge boson vertex functions, $^*\Gamma^{\mu}$, are given by%
\begin{eqnarray}
\label{eq:gamma}
{}^*\Gamma^{\mu}_{\alpha \beta \gamma} & \equiv & \gamma^{\mu}_{\alpha \beta \gamma} \ + \ \delta 
\Gamma^{\mu}_{\alpha \beta \gamma} ,  \\
\gamma_{RAA}^{\mu} &=& \gamma_{AAR}^{\mu} \ = \ \gamma^{\mu} , \ \ \ \mbox{otherwise} \ \ \ 0 . \nonumber 
\end{eqnarray}
where $\delta \Gamma^{\mu}_{\alpha \beta \gamma}$ denotes the HTL resummed contribution to the vertex function.\cite{Braaten}

As mentioned in the introduction, at zero temperature the fermion wave function renormalization constant $A(P)$ is equal 
to unity in the Landau gauge ($\xi=0$) even in the ladder (point-vertex) DS equation, while at finite temperature 
it is not. The quantity $C(P)/A(P) \equiv M(P)$ plays the role of the mass function, in which we are interested, that vanishes 
in the chiral symmetric phase.

\subsection{The HTL resummed DS equations for the invariant functions $A$, $B$ and $C$}
\label{s22}

In the present analysis, we solve the DS equation for the retarded fermion self-energy function $\Sigma_R$, 
with the HTL resummed  gauge boson propagator, by adopting further the following two approximations, 
i) the point-vertex approximation, and ii) the modified instantaneous exchange approximation, on which we give brief 
explanations below.

\vspace{0.3cm}
\begin{flushleft}
i) Point-vertex approximation
\end{flushleft}

As for the vertex function $^*\Gamma^{\mu}$ we adopt the point-vertex approximation, namely we disregard  
$\delta \Gamma^{\mu}_{\alpha \beta \gamma}$ in Eq.~(\ref{eq:gamma}). Thus we investigate the ladder (point-vertex) DS equation 
with the HTL resummed gauge boson propagator.

There are two reasons; Firstly, without the point-vertex approximation the numerical calculation we should carry out becomes so complicated 
that we can not manage with the power of the computer we use, because the HTL resummed contribution to the vertex function, 
$\delta \Gamma^{\mu}_{\alpha \beta \gamma}$, is the non-local interaction term, and also because it behaves singular in numerical 
calculations. Secondly, in the DS equation with the HTL resummed vertex function, it is difficult to resolve the problem of double 
counting of diagrams,\cite{Aurenche} especially at the level of numerical analyses. Being free from this problem is the main reason 
we make use of the point-vertex approximation.

\vspace{0.3cm}
\begin{flushleft}
ii) Modified Instantaneous Exchange (MIE) approximation
\end{flushleft}

We make use of the modified instantaneous exchange (IE) approximation (i.e., set the energy component 
of the gauge boson to be zero) to the gauge boson propagator, which consists of taking the IE limit in the HTL resummed 
longitudinal (electric) gauge boson propagator, $^*G^{\mu \nu}_L$, that is proportional to $B^{\mu \nu}$, while keeping the 
exact HTL resummed form for the transverse (magnetic) gauge boson propagator, $^*G^{\mu \nu}_T$, that is proportional 
to $A^{\mu \nu}$, and also for the massless gauge term in proportion to $D^{\mu \nu}$. The reason why we do not take 
the IE limit to the transverse mode is that the IE approximation reduces the transverse mode to the pure massless propagation, 
thus makes the important thermal effect, i.e., the dynamical screening of transverse propagation disappears.

\vspace{0.2cm}

With the above two approximations, we obtain the HTL resummed improved ladder DS equations for the invariant scalar 
functions $A$, $B$ and $C$:
\begin{eqnarray}
\label{eq:A}
 & & p^2[1-A(P)] = e^2 \left. \int \frac{d^4K}{(2 \pi)^4}
       \right[ \{1+2n_B(p_0-k_0) \} Im[\ ^*G^{\rho \sigma}_R(P-K)]
       \times  \nonumber \\
  & & \Bigl[ \{ K_{\sigma}P_{\rho} + K_{\rho} P_{\sigma}
       - p_0 (K_{\sigma} g_{\rho 0} + K_{\rho} g_{\sigma 0} ) 
       - k_0 (P_{\sigma} g_{\rho 0} + P_{\rho} g_{\sigma 0} )
       + pkz g_{\sigma \rho} \nonumber \\
  & & + 2p_0k_0g_{\sigma 0}g_{\rho 0} \}\frac{A(K)}{[k_0+B(K)+i
       \epsilon]^2 - A(K)^2k^2 -C(K)^2 }
       + \{ P_{\sigma} g_{\rho 0} + P_{\rho} g_{\sigma 0} \nonumber \\
  & & - 2p_0 g_{\sigma 0} g_{\rho 0} \}
       \frac{k_0+B(K)}{[k_0+B(K)+i \epsilon]^2 - A(K)^2k^2
       -C(K)^2 } \Bigr] + \{1-2n_F(k_0) \}
       \times \nonumber \\ 
  & & \ ^*G^{\rho \sigma}_R(P-K) Im \Bigl[
       \{ K_{\sigma}P_{\rho}  + K_{\rho} P_{\sigma} - p_0 (K_{\sigma}
       g_{\rho 0} + K_{\rho} g_{\sigma 0} ) - k_0 (P_{\sigma}
       g_{\rho 0} + P_{\rho} g_{\sigma 0} ) \nonumber \\
  & & + pkz g_{\sigma \rho} + 2p_0k_0g_{\sigma 0}g_{\rho 0}\}
       \frac{A(K)}{[k_0+B(K)+i \epsilon]^2 - A(K)^2k^2-C(K)^2 } 
       \nonumber \\
  & & \left. +  \{ P_{\sigma} g_{\rho 0} + P_{\rho} g_{\sigma 0}
       - 2p_0 g_{\sigma 0} g_{\rho 0} \}
       \frac{k_0+B(K)}{[k_0+B(K)+i \epsilon]^2 - A(K)^2k^2
       -C(K)^2 } \Bigr] \right] \ , \nonumber \\
& & \\
\label{eq:B}
& &  B(P) = e^2 \left. \int \frac{d^4K}{(2 \pi)^4} \right[
        \{1+2n_B(p_0-k_0)\} Im[\ ^*G^{\rho \sigma}_R(P-K)] \times
         \nonumber \\
  & & \Bigl[ \{ K_{\sigma} g_{\rho 0} + K_{\rho} g_{\sigma 0}
       - 2k_0 g_{\sigma 0} g_{\rho 0} \}
       \frac{A(K)}{[k_0+B(K)+i \epsilon]^2 - A(K)^2k^2
       -C(K)^2 } \nonumber \\
  & & + \{ 2g_{\rho 0} 2g_{\sigma 0} - g_{\sigma \rho} \} 
       \frac{k_0+B(K)}{[k_0+B(K)+i \epsilon]^2 - A(K)^2k^2-C(K)^2 }
       \Bigr] + \{1-2n_F(k_0) \} \times \nonumber \\ 
  & & \ ^*G^{\rho \sigma}_R(P-K) Im \Bigl[ \frac{A(K)}{[k_0+B(K)+i
       \epsilon]^2 - A(K)^2k^2 -C(K)^2 } 
       \{ K_{\sigma} g_{\rho 0} + K_{\rho} g_{\sigma 0}  \nonumber \\
  & & \left. - 2k_0 g_{\sigma 0} g_{\rho 0} \} + \frac{k_0+B(K)}{[k_0+B(K)+
       i \epsilon]^2 - A(K)^2k^2-C(K)^2 }
       \{ 2g_{\rho 0} 2g_{\sigma 0} - g_{\sigma \rho} \} \Bigr] 
       \right] \ , \nonumber \\
& & \\
\label{eq:C}
& &  C(P) = -e^2 \int \frac{d^4K}{(2 \pi)^4} g_{\sigma \rho} 
       \{1+2n_B(p_0-k_0) \} Im[\ ^*G^{\rho \sigma}_R(P-K)]
       \times \nonumber \\
  & & \Bigl[ \frac{C(K)}{[k_0+B(K)+i \epsilon]^2 - A(K)^2k^2
       -C(K)^2 } + \{1-2n_F(k_0) \} \times \nonumber \\ 
  & & \left. \ ^*G^{\rho \sigma}_R(P-K) Im \Bigl[
       \frac{C(K)}{[k_0+B(K)+i \epsilon]^2 - A(K)^2k^2
       -C(K)^2 } \Bigr] \right] \ ,
\end{eqnarray}

The HTL resummed DS equations are the coupled integral equations for the six unknown functions because the invariants 
$A$, $B$ and $C$ have both real and  imaginary parts. Therefore, the DS equations, Eqs.~(\ref{eq:A})-(\ref{eq:C}), 
are still quite tough to be solved even if we adopt the above two approximations.

\subsection{The effective potential $V[S_R]$ for the retarded full fermion propagator $S_R$}

The above DS equations, Eqs.~(\ref{eq:A})-(\ref{eq:C}), may have several solutions, and we choose the ``true'' solution by 
evaluating the effective potential $V[S_R]$ for the fermion propagator function $S_R$, then finding the lowest energy 
solution. The effective potential is expressed as\cite{CJT}
\begin{eqnarray}
\label{eq:potential}
 V [ S_R ] \!\! & =& \!\! i \mbox{Tr} \left[ P\!\!\!\!/ S_R \right]  +  i \mbox{Tr} \ln  \left[  i S_R^{-1} \right]  \nonumber \\
       & & - \frac{e^2}{2} \int \frac{d^4K}{(2 \pi)^4} \int \frac{d^4P}{(2 \pi)^4}
             \frac12 \mbox{tr} \left[ \gamma_{\mu} S_R(K) \gamma_{\nu} S_R(P) D_C^{\mu \nu} (P-K) \right. \nonumber \\
       & & \ \ \ \ \ \ \left.
               +  \gamma_{\mu} S_C(K) \gamma_{\nu} S_R(P) D_R^{\mu \nu} (P-K)
               +  \gamma_{\mu} S_R(K) \gamma_{\nu} S_C(P) D_A^{\mu \nu} (P-K) \right] , \nonumber \\
& & 
\end{eqnarray}
where the 1st and the 2nd terms correspond to the one-loop contribution, while the 3rd term corresponds to the two-loop contribution.

\section{Solution of the HTL improved ladder DS equation and its gauge-dependence} 
\label{s3}

In this section we solve the HTL improved ladder DS equation derived in the previous section numerically\footnote{As is 
evident from the definition, Eq.~(\ref{eq_sigma}), the invariants $B$ and $C$ have a dimension of mass. Thus, in 
this section, they are measured in units of the cutoff scale $\Lambda$ in all the results shown.} by an iterative method.
To start with we choose appropriate initial trial functions for $A(P)$, $B(P)$ and $C(P)$ with the guess for them to  have 
non-trivial imaginary parts. By calculating the right-hand sides of Eqs.~(\ref{eq:A})-(\ref{eq:C}) we get ``solutions'' 
$A(P)$, $B(P)$ and $C(P)$, which are supposed to be better approximations to the real solutions.  Replacing the initial 
trial functions for $A(P)$, $B(P)$ and $C(P)$ by the ``solutions'' thus got, we obtain further better approximations to 
the solutions. This procedure is iterated till we obtain converged solutions. Since the solutions thus determined may 
depend on the initial choice of the trial functions, we need to try various initial trial functions to find the 
real solutions. Some technical details in the numerical analysis in solving the DS equations are given in the Appendix.

If more than two converged solutions are obtained, among the solutions we adopt such a solution to be the true one that has 
the lowest value of the effective potential $V[S_R]$, Eq.~(\ref{eq:potential}).

To investigate how the solution thus obtained depends on the choice of gauges, we solve
the HTL improved ladder DS equation by choosing various gauges within the covariant gauge. In the present paper, we carry 
out the analysis by choosing gauges  mainly in the neighborhood of the Landau gauge ($\xi =0$), and show explicitly the obtained 
solutions suffer from the strong gauge-dependence. 

Gauges we choose are those with five different values of the gauge-parameter $\xi$, including the Landau gauge: 

$\xi=0$(Landau gauge), $\xi= \pm 0.05$ and $\xi= \pm 0.025$.

\begin{figure}[tb]
\centerline{\includegraphics[width=12cm]{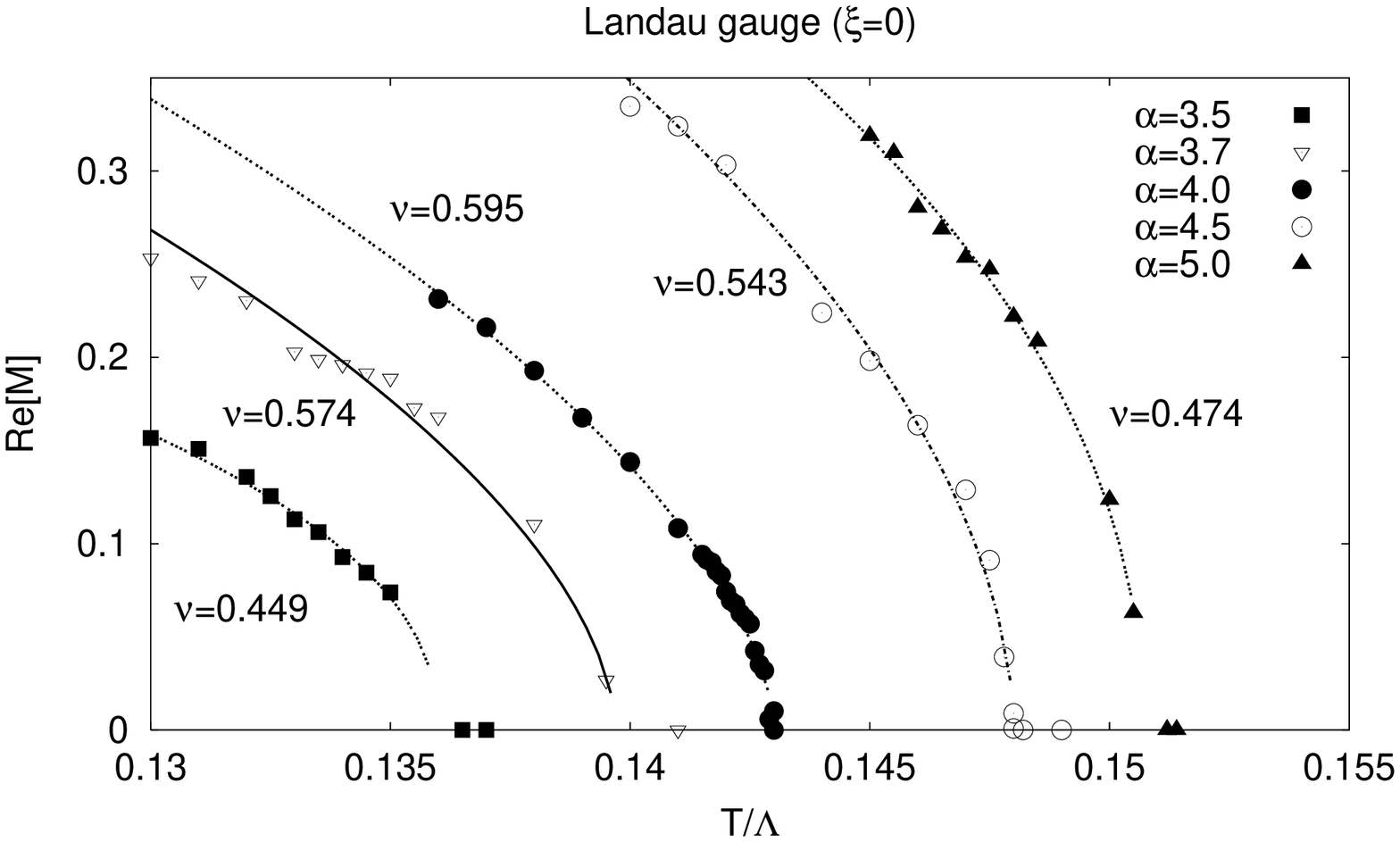}}
\caption{The $T$-dependence of the mass $Re[M]=Re[C/A]$ at $p_0=0$, $p=0.1  \Lambda$ for various fixed values of the coupling 
constant $\alpha$ in the Landau gauge $(\xi=0)$. The best-fit curves at each coupling constant, with the critical 
exponents $\nu$ given in each data, are also shown.}
\label{fig_1}
\centerline{\includegraphics[width=12cm]{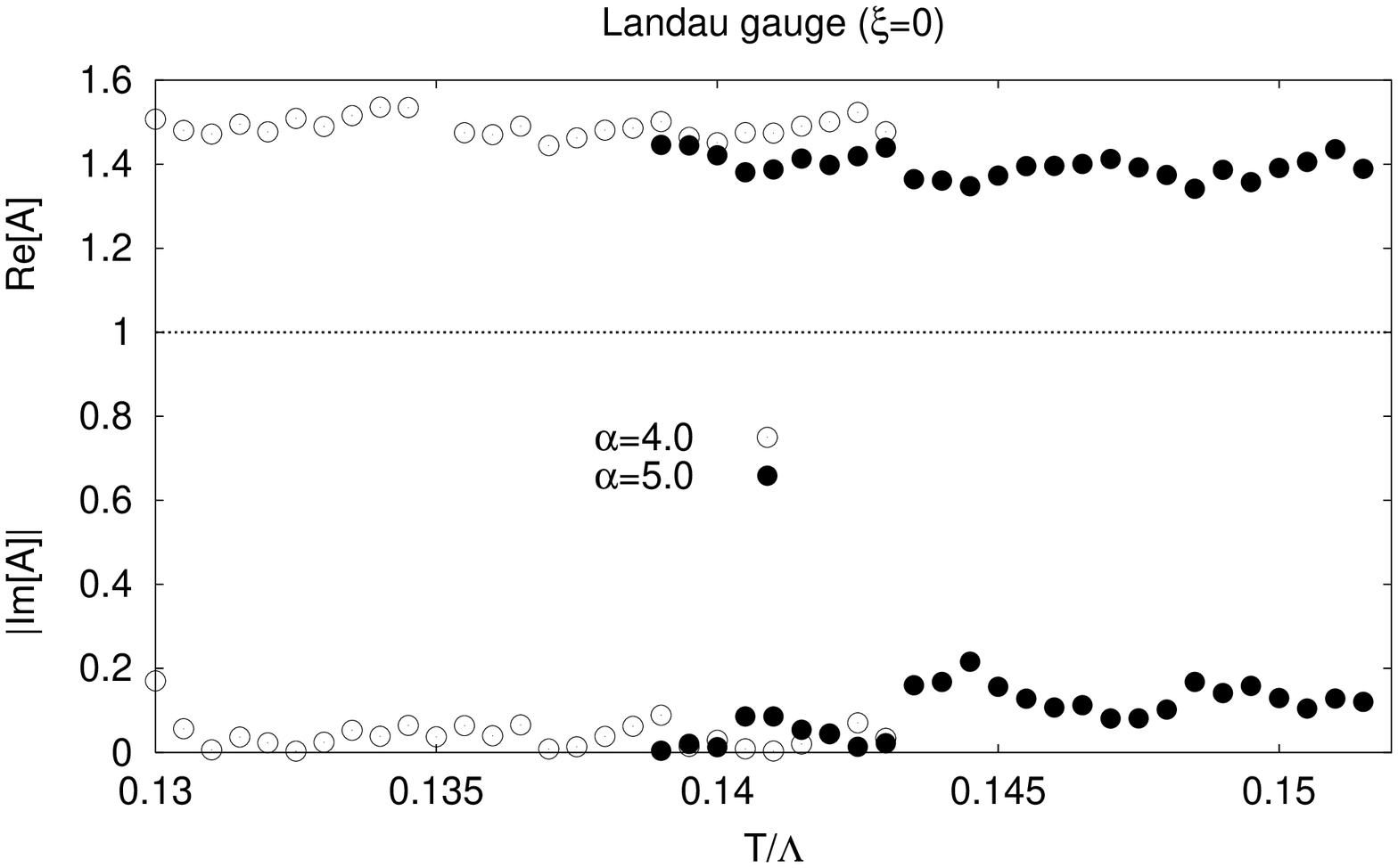}}
\caption{Comparison of the wave function renormalization constant $Re[A]$ and $|Im[A]|$ at the coupling constant
$\alpha =4.0$ and $5.0$ evaluated at $p_0=0$, 
$p=0.1  \Lambda$ in the Landau gauge $(\xi=0)$.}
\label{fig_2}
\end{figure}

\subsection{ Result in the Landau gauge ($\xi=0$)}
\label{s31}

Firstly we give the result in the Landau gauge, which is the gauge studied in most of the preceding analyses. Fig.~1 shows the mass 
$Re[M(P)] \equiv Re[C(P)/A(P)]$ as a function of temperature $T$ at five different values of the coupling constant $\alpha$, 
$\alpha=3.5$, $3.7$, $4.0$, $4.5$ and $5.0$.
In this figure we also give the critical exponent $\nu$ at each coupling constant $\alpha$ defined by
\begin{equation}
\label{nu}
 Re[M] = C_{\alpha} \left( T_c - T \right)^{\nu}, \ \ \ T < T_c,
\end{equation}
which controls how the mass $Re[M]$ vanishes near the  critical temperature $T_c$. We see that the temperature dependence of
the mass $Re[M]$ near the critical temperature $T_c$ can be well described by the functional form Eq.~(\ref{nu}) with little
depnedence of the critical exponent on the couplng constant. The average value $\nu \simeq 0.527$ can reproduce well
the result with appropriate $C_{\alpha}$.
Thus we may say that the phase transition takes place through the second order transition.\footnote{To be exact,
in order to conclude that the phase transition is of the second order, we must study at each coupling constant whether 
there are no stable states other than the one shown in the figure.}

However, this result can not be justified  in having a physical significance without further consideration. As shown in Fig.~2, 
the invariant function $A(P)$ at the momentum where the fermion mass is defined deviates largely from 1, even its imaginary part 
being sizable: $Re[A] \gtrsim 1.4$ and $|Im[A]| \simeq 0.1 \sim 0.2$, indicating $Z_2$ significantly 
smaller than 1, not even the real number. This fact implies that the result obtained in the Landau gauge apparently contradicts 
with the Ward-Takahashi identity $Z_1=Z_2$, therefore we should not give a serious physical meaning to the result in the 
Landau gauge.

\begin{figure}[tbp]
\centerline{\includegraphics[width=12cm]{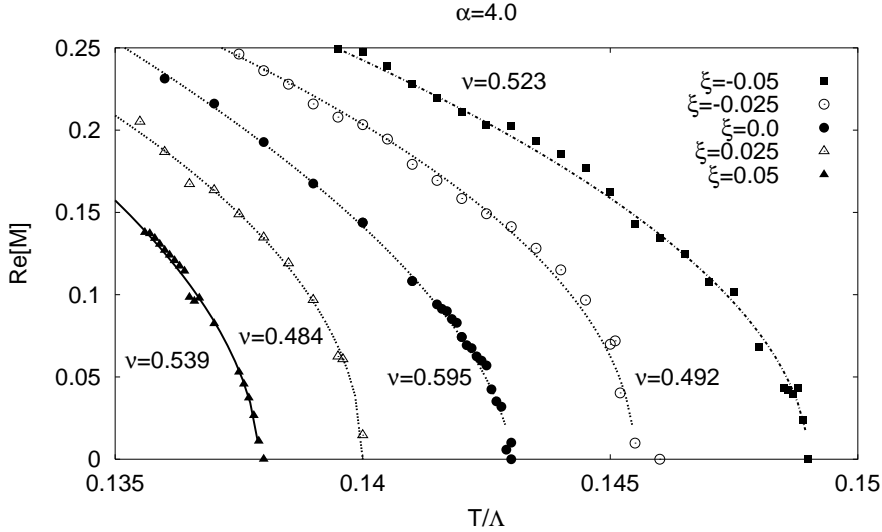}}
\caption{Gauge-parameter-dependence of the fermion mass $Re[M]=Re[C/A]$ at the coupling constant $\alpha =4.0$ evaluated
at $p_0=0$, $p=0.1 \Lambda$. The best-fit curves for each value of the gauge-parameter, with the critical 
exponents $\nu$ given in each data, are also shown.}
\label{fig_3}
\end{figure}

\subsection{Gauge-dependence of the solution}
\label{s32}

Now we compare the results with the five different values of $\xi$ in order to see how large the solution 
depends on the choice of gauge.  We illustrate it by showing the results obtained in the case of the coupling 
constant $\alpha=4.0$. Fig.~3 shows how the mass $M(P)$, as a function of temperature $T$, depends on the 
choice of gauge-parameters. 

\begin{figure}[tbp] 
\centerline{\includegraphics[width=12cm]{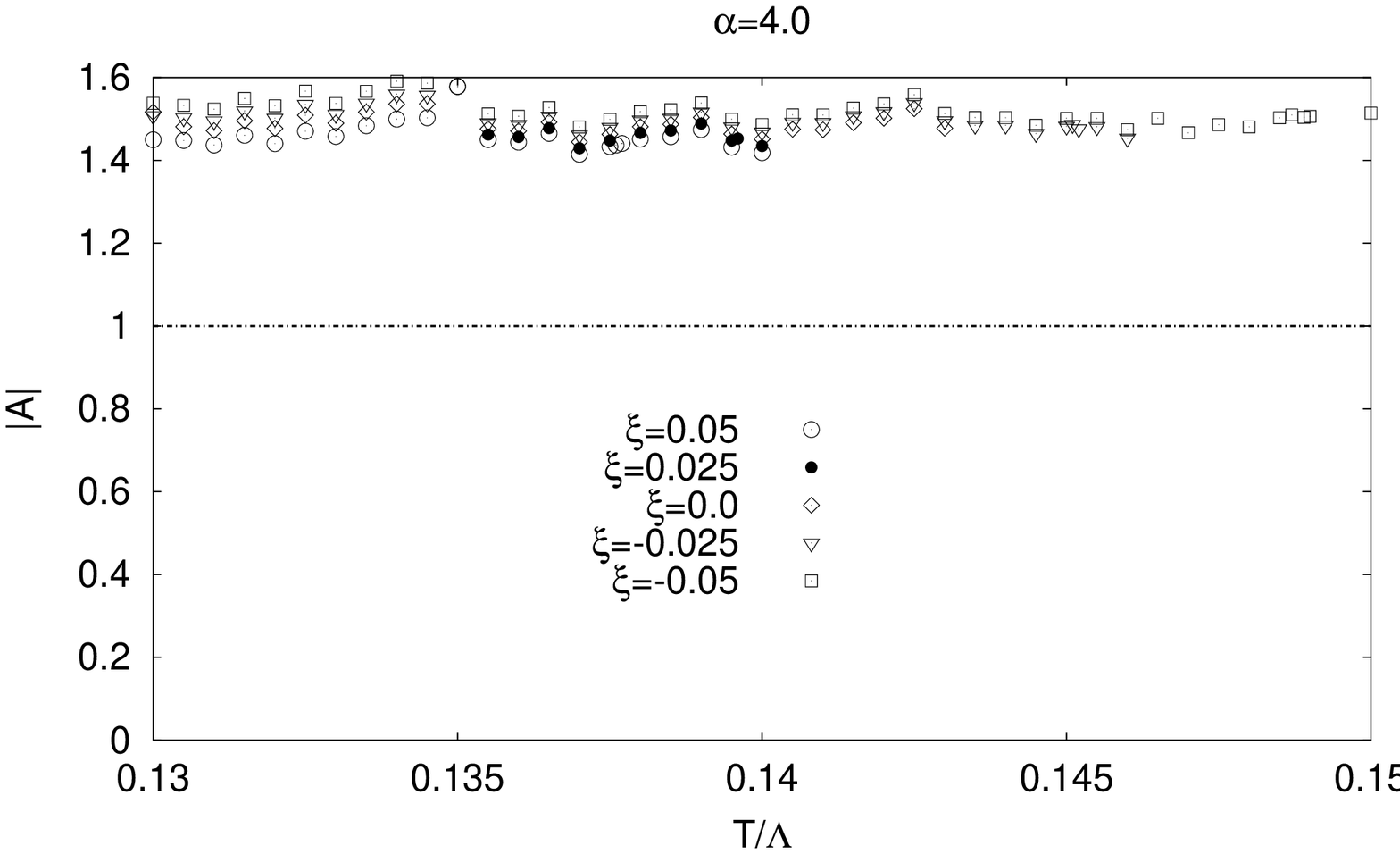}}
\caption{Comparison of the wave function renormalization constant $|A|$ at the coupling constant $\alpha =4.0$ 
evaluated at $p_0=0$, $p=0.1  \Lambda$.}
\label{fig_4}
\centerline{\includegraphics[width=12cm]{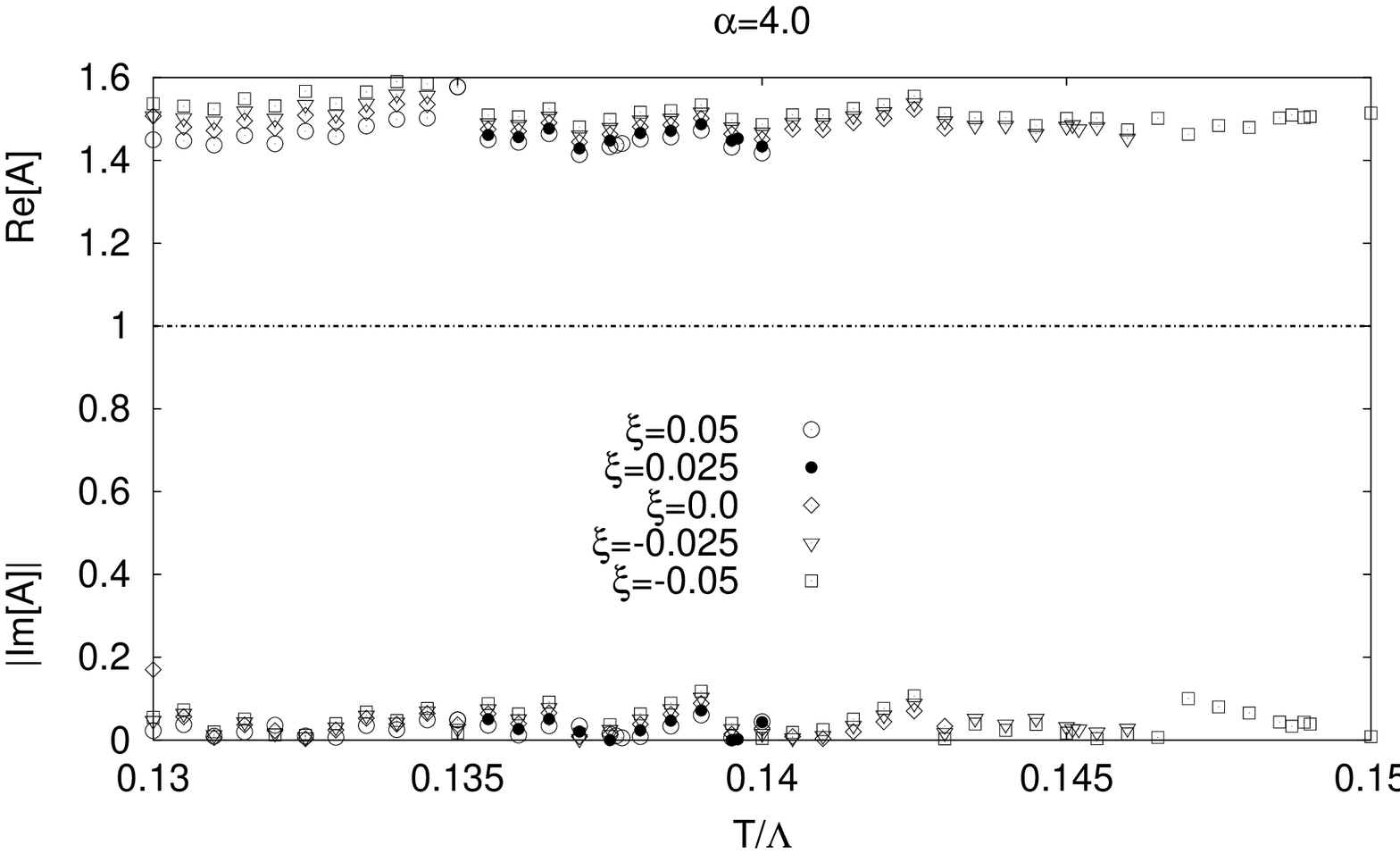}}
\caption{Comparison of the wave function renormalization constant $Re[A]$ and $|Im[A]|$ at the coupling 
constant $\alpha =4.0$ evaluated at $p_0=0$, $p=0.1  \Lambda$.}
\label{fig_5}
\end{figure}

In this figure we give also the critical exponent $\nu$ at each value of the gauge-parameter. In any gauge the 
temperature-dependence of the mass $Re[M]$ near the critical temperature can be well described by the 
functional form Eq.~(\ref{nu}) with little depnedence of the critical exponent on the couplng constant. 
The average value $\nu \simeq 0.527$ can reproduce well the result with appropriate $C_{\alpha}$.
This fact indicates that the phase transition takes place through the second order transition.

However, as can be easily seen, the critical temperature strongly depends on the choice of gauge-parameter, 
demonstrating that it is essential to choose an appropriate gauge such that the Ward-Takahashi identity holds 
in solving the improved ladder DS equation in order to obtain reliable results which have predictive power.

For further consideration of the optimal gauge, we show the gauge dependence of the wave-function renormalization 
function $A(P)$ in Figs.~4 and 5. 
Fig.~4 shows the absolute value of $A(P)$ as a function of temperature for the five gauge parameters, and Fig.~5 
shows the real and imaginary parts of $A(P)$. 
As can be seen from these figures the Landau gauge does not give any better property compared with other gauge 
parameters: obviously the Ward-Takahashi identity does not favor the Landau gauge at all !

\section{Conclusions and discussion}
\label{s4}
In the present paper we solved with a numerical procedure the HTL improved ladder DS equation for the retarded 
fermion self-energy function $\Sigma_R$ to study the spontaneous generation of fermion mass in thermal QCD/QED, 
mainly focussing on the gauge-dependence of the solution within a general covariant gauge where the gauge
parameter $\xi$ is any constant number. It should be noticed that, in the DS equation in the point-vertex 
ladder approximation, no solution receives the vertex correction, thus the vertex renormalization constant $Z_1$ is 
exactly unity, $Z_1=1$. We also made an improvement in estimating the numerical integration over singular parts of 
the integration kernel in the present analysis.

With the numerical solutions thus obtained in various gauges, we found the followings;
\begin{list}{\roman{beanr})}{\usecounter{beanr}}
  \item In the Landau gauge the obtained solution shows a significant change compared to the simple ladder 
        analyses,\cite{Barducci} indicating the importance of taking the dominant effect of thermal fluctuations 
        into the integration kernel through the HTL resummation procedure. 
  \item In any gauge (including the Landau gauge $\xi=0$) where the gauge parameter $\xi$ is any constant number, 
        the fermion wave function renormalization function $A(P)$ always deviates largely from unity even at the momentum 
        where the mass is defined. This fact clearly shows that the corresponding solutions explicitly contradict with 
        the Ward-Takahashi identity  $Z_1=Z_2$, which makes the physical meaning of the solution being obscure.
  \item The obtained solutions strongly depend on the choice of gauge parameters: the critical temperatures (and the 
        critical coupling constants) change significantly gauge by gauge. 
  \item We also determined the critical exponent $\nu$ defined by Eq.~(\ref{nu}), which controls how the mass $Re[M]$ vanishes 
        near the critical temperature $T_c$. The results show that the temperature-dependence of mass  $Re[M]$ near the critical 
        temperature $T_c$ can be well described by the functional form Eq.~(\ref{nu}), and that $\nu$ does not depend 
        significantly on the strength of the coupling, nor on the choice of gauge. Thus we may conclude that, despite the 
        fact iii), the order of phase transition is consistent with the second order transition.
\end{list}

All the above findings show the solution of the  HTL improved ladder DS equation suffers from the problem of large 
gauge-dependence within a general covariant gauge where the gauge parameter $\xi$ is any constant number. Namely the 
solution varies significantly gauge by gauge. The most serious problem we face is there is no definite criterion which 
solution we should choose, which then reminds us of the fact repeatedly we have mentioned that at zero temperature any 
solution in the Landau gauge of the DS equation with the ladder kernel automatically satisfies the Ward-Takahashi identity
$Z_1=Z_2$: one of the most promising criterion selecting the solution to have definite physical meaning.

Here we give some comment on the choice of the gauge in the present analysis. As we noted in $\S$\ref{s3}, we solved the HTL
improved ladder DS equation by choosing gauges only in the neighborhood of the Landau gauge ($\xi =0$). The reason why we choose
such gauges is as follows; i) The Landau gauge has a special significance at zero temperature, and might do so even at finite 
temperatures. With this expectation many analyses have been carried out in the Landau gauge, thus in performing the analysis 
in the neighborhood of the Landau gauge, we can see what  really happens by comparing the result of our analysis with those 
of the preceding works. ii) There is a more practical reason: in our present procedure we can get nicely converged numerical 
solutions mainly in the gauges neighboring with the Landau gauge. 

Anyway with such a small change of gauge, the solution we obtained shows a big change in the ``physical quantities'' such as
the critical temperature.  

Thus the only conclusion we could have from our analysis is that the chiral phase transition in massless thermal QED/QCD at zero 
fermion number density takes place through the second order transition. To determine the critical temperature, the critical 
coupling constant and also the corresponding critical exponents in a physically sensible way, we should find such a solution 
that satisfies at least the Ward-Takahashi identity  $Z_1=Z_2$.

Needless to say, any solution of the DS equation in the ladder approximation can not satisfy the full Ward-Takahashi identity, 
stating the identity between the vertex function and the derivative of the fermion self-energy function. We only propose a possible 
choice of gauge where  $Z_1=Z_2$ holds at least, which may help us to get a physically sensible solution at the same level of 
significance as that at zero temperature analysis in the Landau gauge. It is our next plan of analysis to carry out this procedure 
and investigate the properties of such solutions.\cite{Nagoya,NYY}

\section*{Acknowledgements}

This work has been supported in part by the Grant-in-Aid for Scientific Research [(C) No.17540271] from the Ministry of Education, 
Culture, Sports, Science and Technology, Japan. 

\appendix
\section{Details of the numerical analysis of the DS equation}
In this Appendix, firstly we explain the problems in the numerical analysis, which we face in solving numerically 
the DS equations, Eqs.~(\ref{eq:A})-(\ref{eq:C}), for the invariant functions $A$, $B$ and $C$, then give the procedures 
we make use of to resolve the problems. We face two problems in the numerical analysis; 
\begin{list}{\roman{beanr})}{\usecounter{beanr}}
 \item As can be seen from Eq.~(\ref{eq:A}), to determine the function $A$, firstly we perform the integration
   over the variable $K$ in the right-hand-side (r.h.s.) of the equation, then divide the result by $p^2$. Needless 
   to say, the $p$-dependences of both sides of the equation should agree with each other, and as it is easy to confirm
   that no problem 
   appears in the analytical calculation. However, a problem does appear in performing the numerical calculation. 
   In the small-$p$ region, we are forced to carry out the numerical integration in the r.h.s. in a higher accuracy 
   level compared with other region of the momentum $p$. This is in fact a hard task, causing larger errors and thus 
   being the origin of the unstable behavior in the numerically determined $A$ in the small-$p$ region. This problem becomes       especially serious in the contribution coming from the term that depends explicitly on the
   gauge-parameter $\xi$, i.e., the $D^{\mu \nu}$ term in the gauge boson propagator ${}^* G^{\mu \nu}$, Eq.~(\ref{eq:Gr}).
   Formally we must divide by $p^3$, not by $p^2$, to determine this contribution to the function $A$.
 \item There are several    singular terms in the gauge boson propagator ${}^* G^{\mu \nu}$, Eq.~(\ref{eq:Gr}), appearing
   in the integration kernel of the DS equations; a) The $\xi$-dependent $D^{\mu \nu}$ term is a pure massless 
   double-pole mode. b) The transverse (magnetic) mode being proportional to $B^{\mu \nu}$ receives the so-called
   dynamical screening, but it becomes massless when the energy-component of the momentum vanishes with the 
   space-component of it being finite. 
\end{list}
   In both cases we face the numerical integration over singular functions, e.g., the principal part and 
   the $\delta$-function. In the analytical calculation these functions do not cause any trouble, but in the numerical
   calculation they, especially the principal part, do cause troubles to obtain stable solutions.
   The procedures we make use of in order to resolve the above problems are as follows;
\begin{list}{\roman{beanr})}{\usecounter{beanr}}
  \item As for the first problem i) above, we carry out the integration in the r.h.s. of Eq.~(\ref{eq:A}) 
    by two different methods depending on the region of the momentum $p$. 
    In the small-$p$ region $p < p_{th}$, as for the contribution coming from the $D^{\mu \nu}$ term that
    depends explicitly on the gauge-parameter $\xi$, we expand the corresponding integrand of the r.h.s. of 
    Eq.~(\ref{eq:A}) in the power series of $p$, keeping up to the $p^3$ term, then carry out the integration. 
    In the large-$p$ region $p > p_{th}$, we perform the ordinary numerical integration to get the function $A$.
       The explicit value of the ``threshold momentum'' $p_{th}$ is determined by considering the stability as 
   well as the smoothness of the solution. In the present analysis we choose $p_{th} = 0.2$. 
 \item As for the second problem, we use the ordinary procedure. When the integration over the principal part 
   appears, we divide the integration region into two parts: the integration in the neighborhood of the singular 
   point, and the integration away from the singular point. We can carry out the simple numerical integration away 
   from the singular point. The integration in the neighborhood of the singular point is carried out analytically, 
   by taking the unknown functions $A$, $B$, and $C$ kept constants with the values of those at the singular point of 
   the gauge boson propagator. 

  In order for this method to work, the unknown functions $A$, $B$ and $C$ should behave smooth in the neighborhood 
   of the singular point of the gauge boson propagator, and also we should take the neighborhood of the singular 
   point as narrow as possible. The first point can be checked a posteriori by the obtained solutions, and the 
   result is satisfactory. As for the size of the neighborhood where the integration is performed analytically, 
   we must choose by seeing the stability and the smoothness of the obtained solutions. In the present analysis, 
   we choose $|p_0 - k_0|< 0.04$ and $|{\bf p} - {\bf k}|< 0.2$.
\end{list}

%

\end{document}